\documentclass{agujournal}
\journalname{JGR-Space Physics}
\graphicspath{ {Figures/}}
\begin{document}

\title{On the statistics of acceleration and loss of relativistic electrons in the outer radiation belt: a superposed epoch analysis}

\authors{Ch. Katsavrias\affil{1,2}, I.A. Daglis\affil{1,2,3} and W. Li\affil{4}}

\affiliation{1}{Department of Physics, National and Kapodistrian University of Athens, Greece.}
\affiliation{2}{Institute of Accelerating Systems and Applications, National and Kapodistrian University of Athens, Greece.}
\affiliation{3}{Institute of Astronomy, Astrophysics, Space Applications and Remote Sensing, National Observatory of Athens, Greece.}
\affiliation{4}{Center for Space Physics, Boston University, Boston, USA.}

\correspondingauthor{Ch. Katsavrias}{ckatsavrias@phys.uoa.gr}

%  List up to three key points (at least one is required)
%  Key Points summarize the main points and conclusions of the article
%  Each must be 100 characters or less with no special characters or punctuation 

\begin{keypoints}
\item The number of events with PSD enhancements is L and $\mu$--dependent.

\item Loss, as a combined effect of magnetopause shadowing and outward diffusion, is a common feature in both depletion and enhancement events.

\item The synergy of enhanced seed population and chorus activity is what distinguishes relativistic electrons enhancements from depletions.
\end{keypoints}

\begin{abstract}
We investigate the response of the outer Van Allen belt electrons to various types of solar wind and magnetospheric disturbances. We use electron phase space density (PSD) calculations as well as concurrent Pc5 and chorus wave activity observations in the outer belt during the Van Allen Probes era to compare 20 electron enhancement and 8 depletion events. Results indicate that the combined effect of magnetopause shadowing and outward diffusion driven by Pc5 waves is present in both groups of events. Furthermore, in the case of enhancement events, the synergy of enhanced seed population levels and chorus activity -- due to the enhanced substorm activity -- can effectively replenish the losses of relativistic electrons, while inward diffusion can further accelerate them.
\end{abstract}
% ------------------------------------------------------------------------ %
%  TEXT
% ------------------------------------------------------------------------ %
%% Suggested section heads:
% \section{Introduction}
% The main text should start with an introduction. Except for short
% manuscripts (such as comments and replies), the text should be divided into sections, each with its own heading. 
% Headings should be sentence fragments and do not begin with a  lowercase letter or number. Examples of good headings are:
% \section{Materials and Methods} % Here is text on Materials and Methods.
% \subsection{A descriptive heading about methods} % More about Methods.
% \section{Data} (Or section title might be a descriptive heading about data)
% \section{Results} (Or section title might be a descriptive heading about the results)
% \section{Conclusions}
% ------------------------------------------------------------------------ %

\section{Introduction}

Earth's outer radiation belt response to geospace disturbances is extremely variable due to the interplay of acceleration and loss mechanisms \citep{Reeves2016a}. Acceleration mechanisms of electrons include inward radial diffusion, which increases the particle energy due to the conservation of the first adiabatic invariant \citep{Schulz1974,Taylor2004,Shprits2008a}, and local acceleration via gyro--resonant interactions with whistler chorus mode waves \citep{Meredith2003,Horne2005,Shprits2008b,Thorne2013,Bortnik2016,Li2016} which violate their first and second invariants. Losses, on the other hand, include scattering into the atmospheric loss cones (drift or bounce) via wave--particle interactions with plasmaspheric hiss, EMIC or chorus waves \citep{Shprits2007,Usanova2014,Jaynes2014} and magnetopause shadowing with subsequent enhanced outward radial transport \citep{Kim2008,Turner2012,Kim2014}. 

During these disturbances, the broad energy range electron population of the outer belt, can be enhanced, depleted, or even not affected at all. Numerous works have performed statistical studies, concerning the electron fluxes in the outer belt \citep{Reeves2003,Zhao2013,Turner2015a,Moya2017} or the total electron content \citep{Murphy2018}, in order to determine the effect of geomagnetic storms to radiation belt electron population. Others \citep{Kilpua2015,Shen2017} have argued that this effect depends on the various solar wind drivers (e.g. ICMEs, SIRs, etc). A step further, \citet{Reeves2016b} showed that the net effect of each storm depends on the electron energy and L--shell.

\citet{Turner2013}, using electron phase space density derived from THEMIS data, showed that from the 53 events under investigation, 58\% resulted in relativistic electron PSD enhancement, 17\% in depletion and 25\% resulted in no significant change in the PSD. Furthermore, by comparing two storms in detail (one that resulted in enhancement and one that resulted in depletion of PSD), they indicated that the storm which resulted in enhancement of relativistic electron PSD exhibited more enhanced and broader range chorus wave amplitudes, more prolonged periods of enhanced Pc5 wave activity but fewer EMIC wave events. At the same extent, \citet{Katsavrias2015b} compared two storms to demonstrate that the storm which resulted in enhancement of relativistic electron PSD exhibited similar chorus and Pc5 enhancements both in amplitude and in temporal/spatial range, while the storm which resulted in depletion of relativistic electron PSD exhibited only a prolonged Pc5 activity with absence of any significant chorus activity.

Regardless of the results, all aforementioned studies have used a threshold of Dst (or Sym-H index) as a selection criterion of geospace disturbances. Nevertheless, the possibility that the Dst index is not related to the mechanisms that cause acceleration and loss of the electron population of the outer belt \citep{Borovsky2006}, renders it an inappropriate selection parameter of the studies of electron variability. This is confirmed by recent studies of relatively weak or even non--storm events which were able to cause enhancement  \citep{Schiller2014} or depletion \citep{Katsavrias2015a} of the relativistic electron population in the outer belt.

To that end, this work presents a new approach, concerning statistical studies of geospace disturbances, by combining both storm and non--storm events. To do that we consider as proxy the maximum compression of the magnetopause instead of the traditional minimum of the Dst index. We also use electron phase space density instead of flux or total electron content in order to take into account the different behaviour of the near--equatorial seed, relativistic and ultra--relativistic electron population at different L--shells.

\section{Data Selection and Methodology}\label{Data}

We make use of the high--quality measurements of both the Magnetic Electron Ion Spectrometer--MagEIS \citep{Blake2013}, and the Relativistic Electron Proton Telescope--REPT \citep{Baker2012} on--board the Van Allen Probes. In order to both track particles and to identify regions and times when the adiabatic assumption breaks down (injection events, fast loss events) we convert particle fluxes to Phase Space Density for fixed values of the three adiabatic invariants, using the method described by \citet{Chen2005}. In detail, we calculate PSD for three values of $\mu$, namely 100, 900 and 4200 MeV/G  for seed, relativistic and ultra--relativistic near equatorial electron population with K$\le0.03$ G$^{1/2}$R$_{E}$, at 4 values of L$^{*}$, namely 3.5 (3.25 $\leq L^{*} \leq$ 3.75), 4 (3.75 $\leq L^{*} \leq$ 4.25), 4.5 (4.25 $\leq L^{*} \leq$ 4.75) and 5 (4.75 $\leq L^{*} \leq$ 5.25). All values of the invariants K and $L^{*}$ are obtained from the magnetic ephemeris of ECT suite (https://www.rbsp-ect.lanl.gov/science/DataDirectories.php) which are calculated using the \citet{Tsyganenko2005} magnetospheric field model (TS05). 

In addition, we apply wavelet analysis on the magnetic field measurements (3--minutes resolution) from the fluxgate magnetometers of RBSP \citep{Kletzing2013}, using the method described in \citep{Balasis2013}. This method allows us to quantify the temporal evolution of ULF--Pc5 wave activity  (typically between 2 and 7 mHz frequency range). 

Finally, we use the technique developed by \citet{Li2013} in order to infer lower--band chorus wave amplitudes. This technique gives us the advantage of having a global indicator of chorus wave activity which cannot be obtained by empirical models or direct wave measurements made by near--equatorially orbiting satellites alone, especially in cases where the apogee of the satellite is on the dusk side, where chorus wave occurrence is typically low. 

Supplementary measurements of five-minute averaged values of solar wind speed, dynamic pressure and interplanetary magnetic field mapped at 1 AU as well as geomagnetic indices Sym-H and AL from the NASA/OMNI database\footnote{http://omniweb.gsfc.nasa.gov/} are also used.
%------------------------------------------------------------------------------------
 \begin{figure}[h]
 \centering
 \includegraphics[width=35pc]{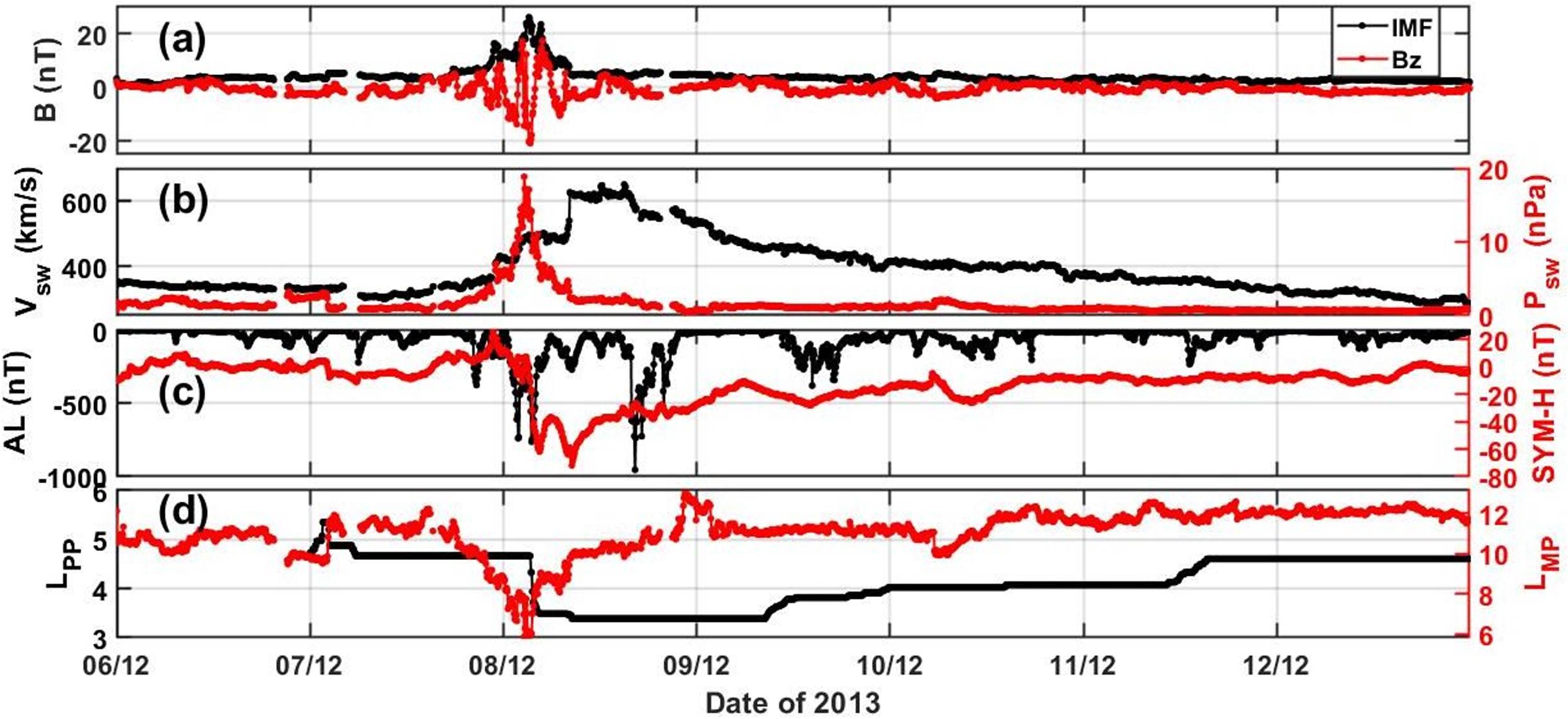}
 \caption{Example of a selected event (December 6--12, 2013) using solar wind and magnetosphere parameters with 5 minutes resolution. Panel a: Interplanetary Magnetic Field (IMF) and its z component ($B_z$); Panel b: Solar wind speed and dynamic pressure; Panel c: The AL and SYM-H geomagnetic indices; Panel d: The location of the dayside magnetopause and the average location of the plasmapause based on the models by \citet{Shue1998} and \citet{OBrien2003a}, respectively.}
 \label{exam}
  \end{figure}
%------------------------------------------------------------------------------------

\section{Event selection and net effect statistics}

We have selected 71 events during the RBSP era, from September 2012 to April 2018, spanning the maximum and declining phase of Solar cycle 24 (see table S1 in supplementary material). For the selection, the following conditions must be satisfied for at least 12 hours before the start of the event: the average solar wind speed and pressure are less than 400 km/s and 3 nPa, respectively, the SYM--H index is continuously higher than -20 nT, the AL index is continuously higher than -300 nT and B$_z$ lies between -5 and 5 nT. The end time of the event is the time when all parameters have return to the above mentioned pre--event levels. An example of such an event is shown in figure \ref{exam}. The values of the aforementioned criteria correspond to the usual variation of solar wind parameters and geomagnetic indices typical during non--storm periods.

In order to determine the net effect of each event we first define as t$_{0}$-epoch the time of the maximum compression of the magnetopause (namely the subsolar standoff distance -- Lmp$_{min}$) as it is given by the model described by \citet{Shue1998}. Then we define as prePSD and postPSD the average PSD during 24 to 48 hours before and 24 to 72 hours after the maximum compression of the magnetopause, respectively. Finally, we divide our sample into 3 categories:

\begin{itemize}
\item Enhancement Events: log$_{10}$(PrePSD) - log$_{10}$(PostPSD) $\leq$ -log$_{10}$(6) 
\item Depletion Events: log$_{10}$(PrePSD) - log$_{10}$(PostPSD) $\geq$ log$_{10}$(4) 
\item No significant change Events: -log$_{10}$(6) $\leq$ log$_{10}$(PrePSD) - log$_{10}$(PostPSD) $\leq$ log$_{10}$(4) 
\end{itemize} 

The log$_{10}$(4) and -log$_{10}$(6) values, which are slightly higher than the ones used in previous studies \citep{Reeves2003,Turner2013,Kilpua2015}, correspond to the average logarithmic depletion or enhancement of the PSD for all events. These criteria allow us to use a sample-dependent threshold but more important, to further examine the mechanisms that produce clear and strong enhancement or depletion events since the use of average PSD (after the maximum compression) and not the maximum allows us to neglect intermittent and short--term variations.  

\begin{figure}[h]
 \centering
 \includegraphics[width=18pc]{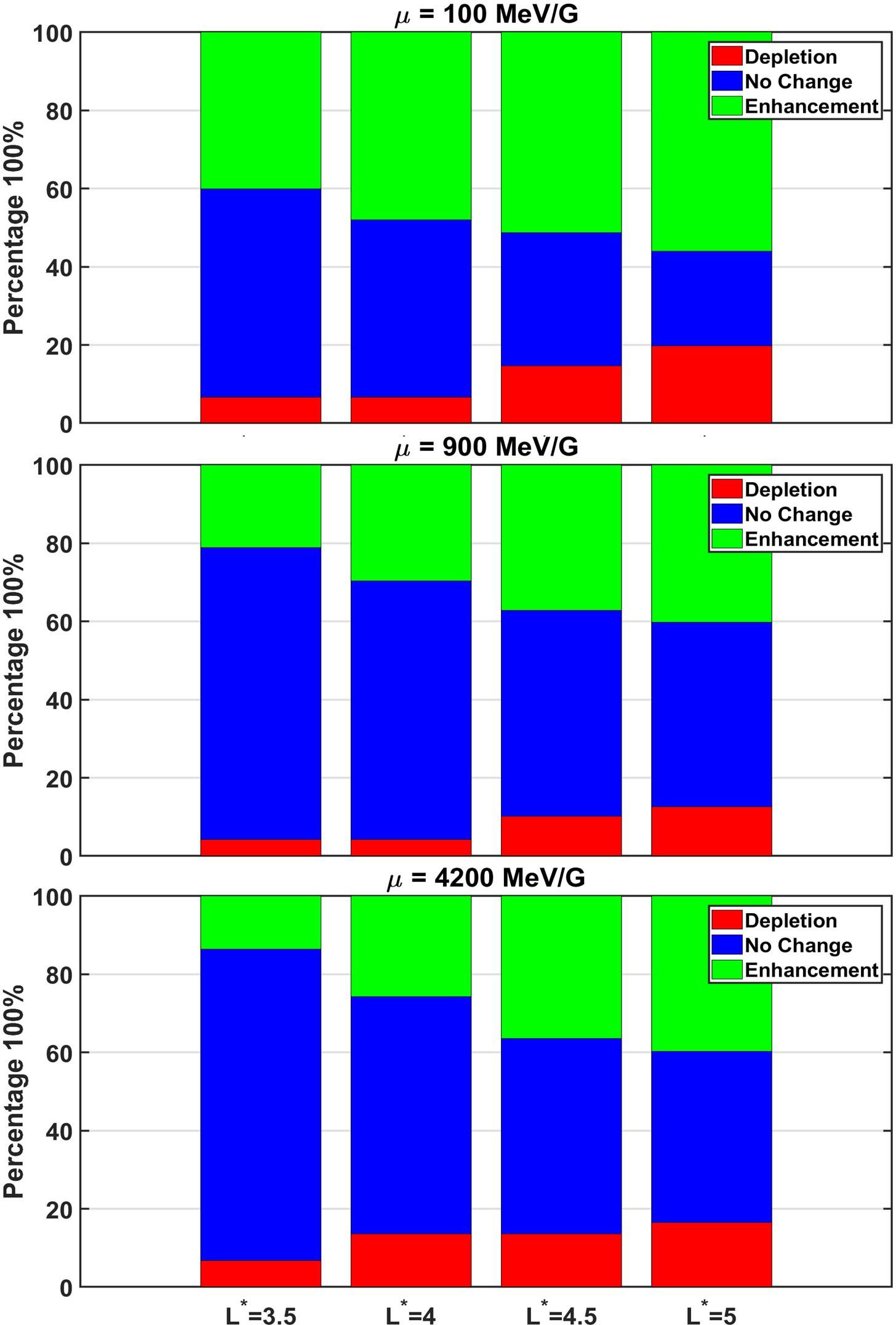}
 \caption{Percentage of event outcome as a function of $\mu$ (100, 900 and 4200 MeV/G) and L$^{*}$ (3.5, 4, 4.5 and 5) for near--equatorial mirroring electrons.}
 \label{dist}
  \end{figure}

Figure \ref{dist}, shows the result of each geospace disturbance concerning the near--equatorial seed, relativistic and ultra--relativistic electron population with K$\le0.03$ G$^{1/2}$R$_{E}$. In order to give an estimation of the energy and pitch angle range of these electrons we use the example of figure \ref{exam}. For this event, the aforementioned value of K corresponds to electrons with equatorial pitch angles in the 70--90 degrees range, while the values of $\mu$ at 3.25$\leq L^{*}\leq$5.25, roughly correspond to an energy range of 0.1--0.6, 0.7--2.4 and 1.9--6.5 MeV for 100, 900 and 4200 MeV/G, respectively. 

As shown, the result of each geospace disturbance is $\mu$- and L$^{*}$-dependent. In detail, concerning the seed population ($\mu$ = 100 MeV/G) at L$^{*}$=5, 57\% of the events resulted in PSD enhancement, 20\% resulted in PSD depletion and 22\% resulted in no significant change. This means that more than half of the selected events were able to introduce significant seed population into the heart of the outer radiation belt. Concerning the relativistic electron population ($\mu$ = 900 MeV/G) at L$^{*}\geq$4.5, 40\% of the events resulted in PSD enhancement, 13\% resulted in PSD depletion and 47\% resulted in no significant change. Concerning, ultra--relativistic population ($\mu$ = 4200 MeV/G) at L$^{*}\geq$4.5, 38\% of the events resulted in PSD enhancement, 17\% resulted in PSD depletion and 45\% resulted in no significant change. With decreasing L$^{*}$, the number of both enhancements and depletions decreases rapidly. 

%------------------------------------------------------------------------------------
%------------------------------------------------------------------------------------

\section{Superposed Epoch Analysis}

We define as enhancement/depletion events those events that resulted in enhancement/depletion, respectively, of the 900 MeV/G electrons at L$^{*}\geq$4.5. From the initial sample of 71 events we have selected 20 enhancement and 8 depletion events which are listed in tables 3 and 4 in the supplementary material. As previously stated, we define as t$_{0}$-epoch the time of the maximum compression of the magnetopause (Lmp$_{min}$) considering as pre--event phase, a period of up to 50 hours before the Lmp$_{min}$ and post--event phase, a period of up to 120 hours after the Lmp$_{min}$. 

%------------------------------------------------------------------------------------
 \begin{figure}[h]
 \centering
 \includegraphics[width=38pc]{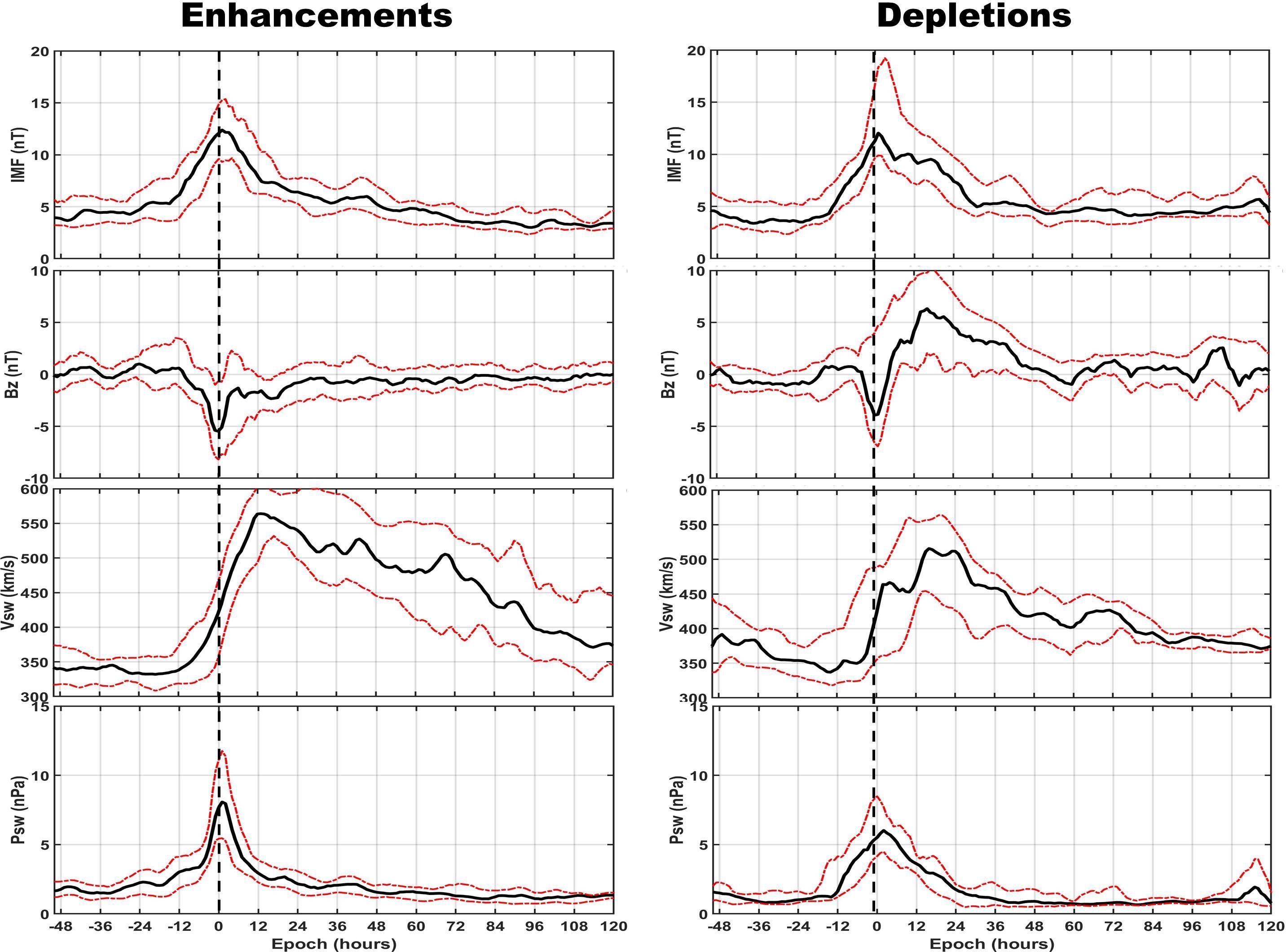}
 \caption{Superposed epoch analysis on solar wind parameter values for enhancement (left) and depletion events (right). The solid black line corresponds to the median, while the dashed red lines to the lower and upper quantiles. The vertical dashed line corresponds to t$_{0}$-epoch which is defined as the maximum compression of the dayside magnetopause. Top to bottom: Interplanetary magnetic field (IMF), its z-component, solar wind speed and dynamic pressure.}
 \label{SW}
  \end{figure}
%------------------------------------------------------------------------------------
Figure \ref{SW} shows the superposed epoch analysis (hence forward SEA) of the solar wind parameters during enhancement (left panels) and depletion events (right panels). All parameters begin to diverge approximately 12 hours before the maximum compression of the magnetopause. Except for IMF, which shows no significant differences between the 2 categories, enhancement events are associated with events characterized by strong pressure pulses, higher and long--lasting values of solar wind speed and continuously negative values of the z--component up to 1.5 day.
%------------------------------------------------------------------------------------
 \begin{figure}[h]
 \centering
 \includegraphics[width=38pc]{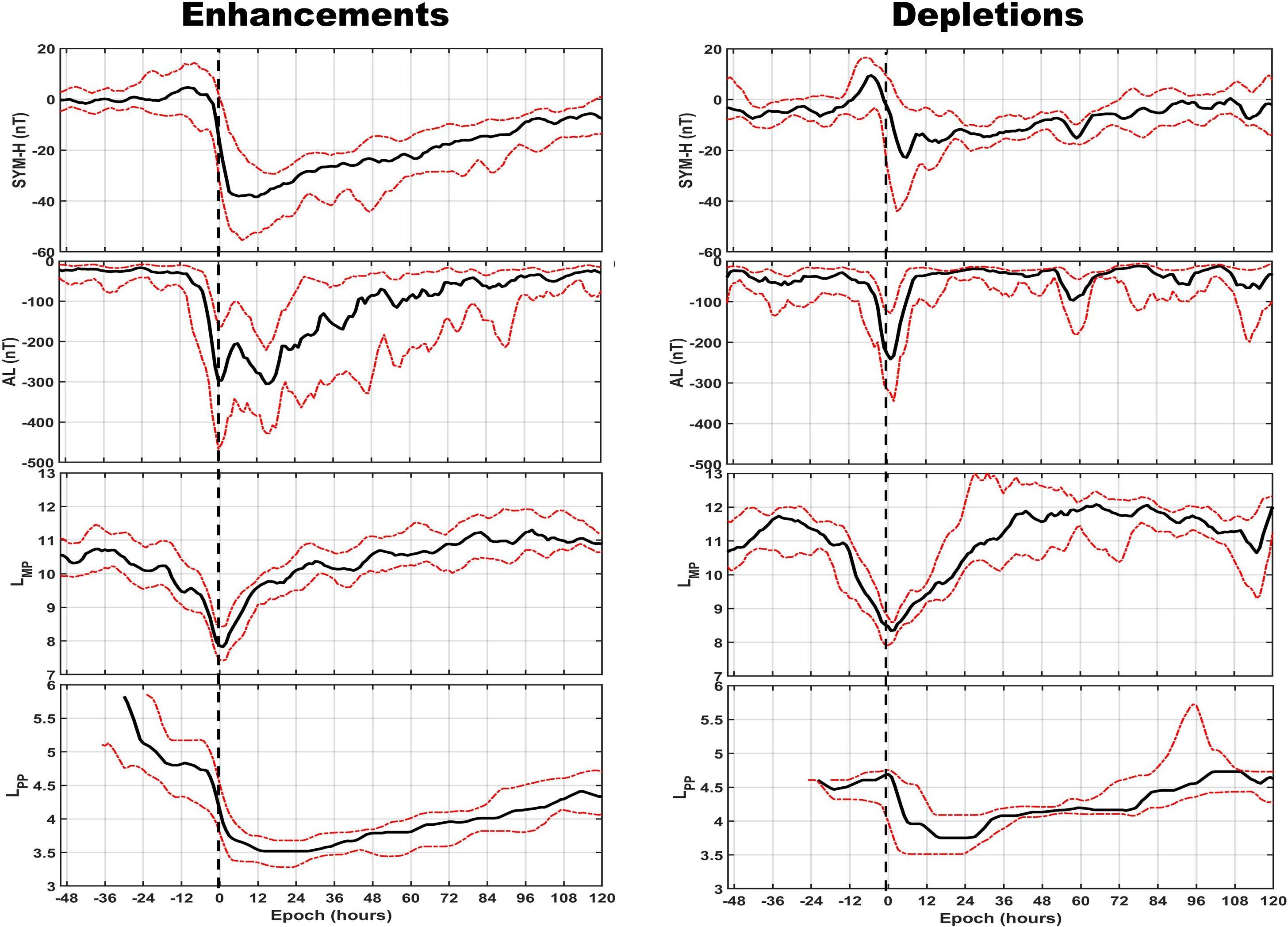}
 \caption{The same as figure \ref{SW} for geomagnetic indices SYM-H and AL, and the location of the magnetopause and plasmapause.}
 \label{GeoMag}
  \end{figure}
%------------------------------------------------------------------------------------
As a natural consequence, events that result in relativistic electron enhancements are also associated with higher levels of magnetospheric activity as reflected in the SYM-H and AL indices (figure \ref{GeoMag}). They are also associated with larger compressions of both the magnetopause and plasmapause. Note, that the most pronounced difference between the two event groups occurs in AL index which is a very good proxy of substorm activity. Enhancement events are characterized by strong and long--lasting (up to 2.5 days) substorm activity. On the contrary, depletion events are connected with weaker and short--lived substorm activity, approximately 6 hours around the maximum compression of the magnetopause. After that, substorm activity can be characterised as negligible. 

%------------------------------------------------------------------------------------
 \begin{figure}[h]
 \centering
 \includegraphics[width=38pc]{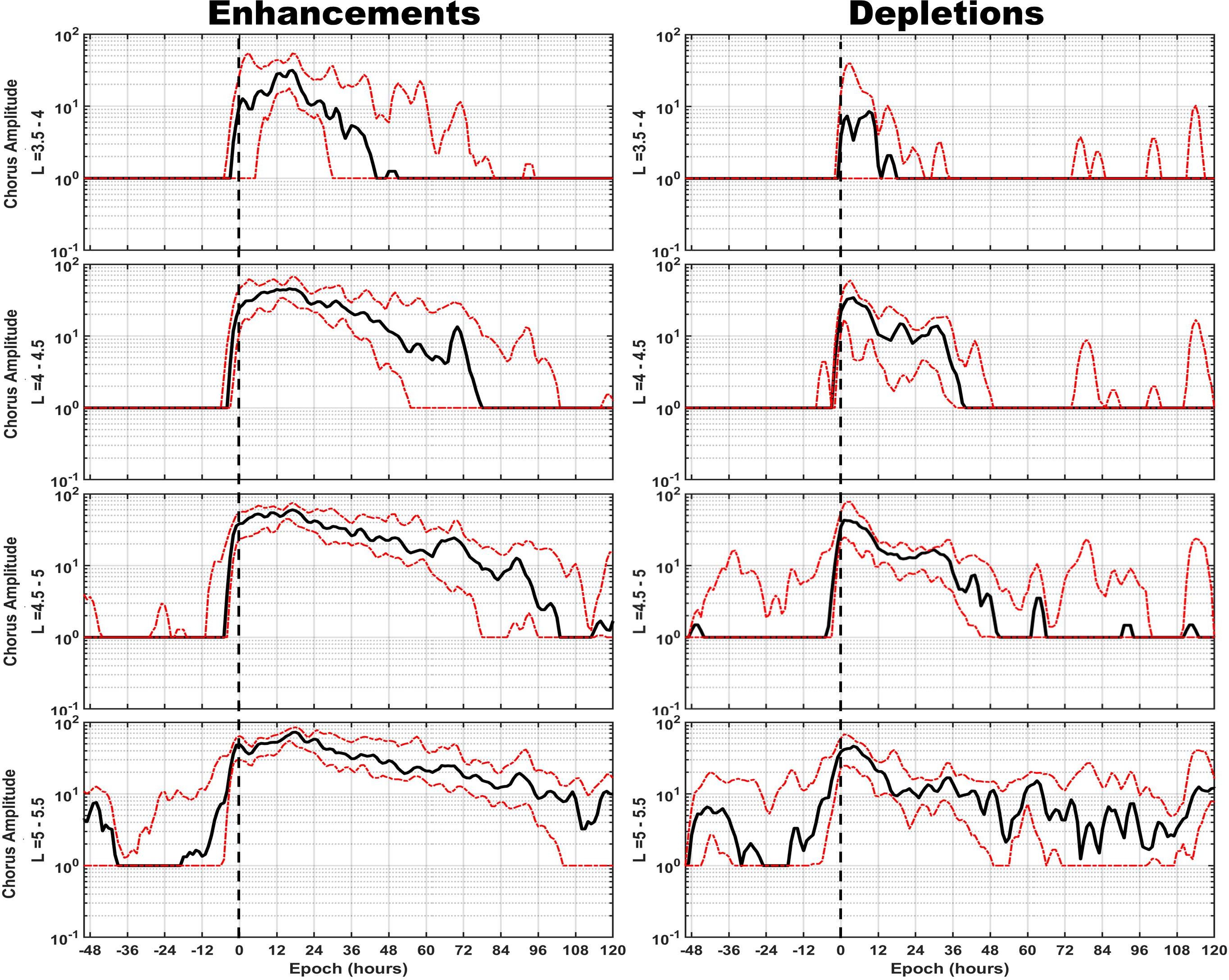}
 \caption{The same as figure \ref{SW} for the normalized chorus amplitude at 4 L--shell ranges: (top to bottom) 3.5<L<4, 4<L<4.5, 4.5<L<5 and 5<L<5.5. Left panels correspond to enhancement events, while right panels correspond to depletion events. }
 \label{CHORUS}
  \end{figure}
%------------------------------------------------------------------------------------
Figure \ref{CHORUS} shows the SEA of the normalized chorus amplitude at 4 L--shell ranges (3.5<L<4, 4<L<4.5, 4.5<L<5 and 5<L<5.5 averaged over a broad range of MLTs inferred by the \citet{Li2013} technique). All values have been normalized to the median between 48 and 24 hours before t$_{0}$--epoch of all enhancement and depletion events, respectively. During enhancement events at 3.5<L<4, chorus amplitude is enhanced right before t$_{0}$-epoch and reaches the maximum (approximately 30 pT) 12 hours later. Moreover, chorus activity remains enhanced up to 36 hours after t$_{0}$-epoch. At higher L--shells, both the maximum amplitude and the duration of chorus activity are increased. Especially at 5<L<5.5, the maximum amplitude has reached approximately 80 pT, while the activity remains above 10 pT for 96 hours after t$_{0}$, continuously. Depletion events exhibit completely different behaviour. Both maximum amplitude and duration are lower than those during enhancement events. At 3.5<L<4, the enhancement is below 1 order of magnitude and chorus remain enhanced only for 12 hours after t$_{0}$, while at 5<L<5.5, the chorus amplitude is above 10 pT up to 36 hours after t$_{0}$.    

%------------------------------------------------------------------------------------
 \begin{figure}[h]
 \centering
 \includegraphics[width=38pc]{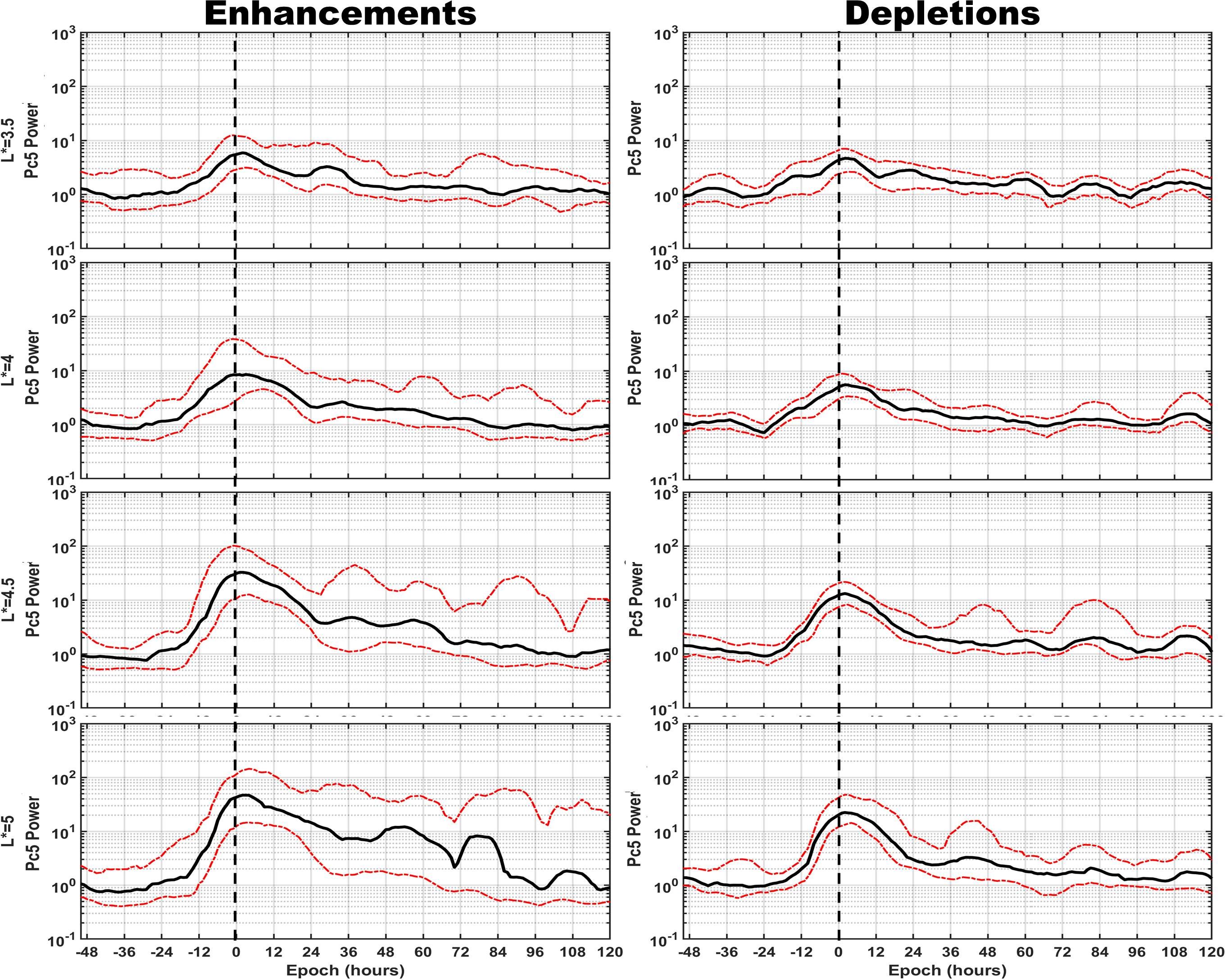}
 \caption{The same as figure \ref{SW} for the normalized average Pc5 power at 4 different values of L$^{*}$: (top to bottom) 3.5, 4, 4.5 and 5.}
 \label{ULF}
  \end{figure}
%------------------------------------------------------------------------------------
Figure \ref{ULF} shows the SEA of the normalized average Pc5 power at 4 different values of L$^{*}$. Both during enhancement and depletion events, Pc5 power at all L$^{*}$, is increased approximately 18--24 hours before t$_{0}$, while the maximum coincides with the maximum of solar wind dynamic pressure, the maximum compression of the magnetopause and the minimum Bz at t$_{0}$. Moreover, the maximum value, for each group of events, is increasing with increasing L$^{*}$. Nevertheless, differences concerning both the maximum power and duration of the Pc5 activity occur between enhancement and depletion events. Enhancements systematically exhibit a slightly larger Pc5 power maximum which is approximately a factor of 10$^{0.3}$ at all L--shells. Concerning the duration, both groups at L$^{*}\leq$4, exhibit enhanced power up to 24--36 hours after t$_{0}$. On the contrary, at L$^{*}\geq$4.5, the enhancement events exhibit enhanced Pc5 power up to approximately 3.5 days after t$_{0}$, while depletion events exhibit enhanced Pc5 power up to approximately 1.5 days.

%------------------------------------------------------------------------------------
 \begin{figure}[h]
 \centering
 \includegraphics[width=38pc]{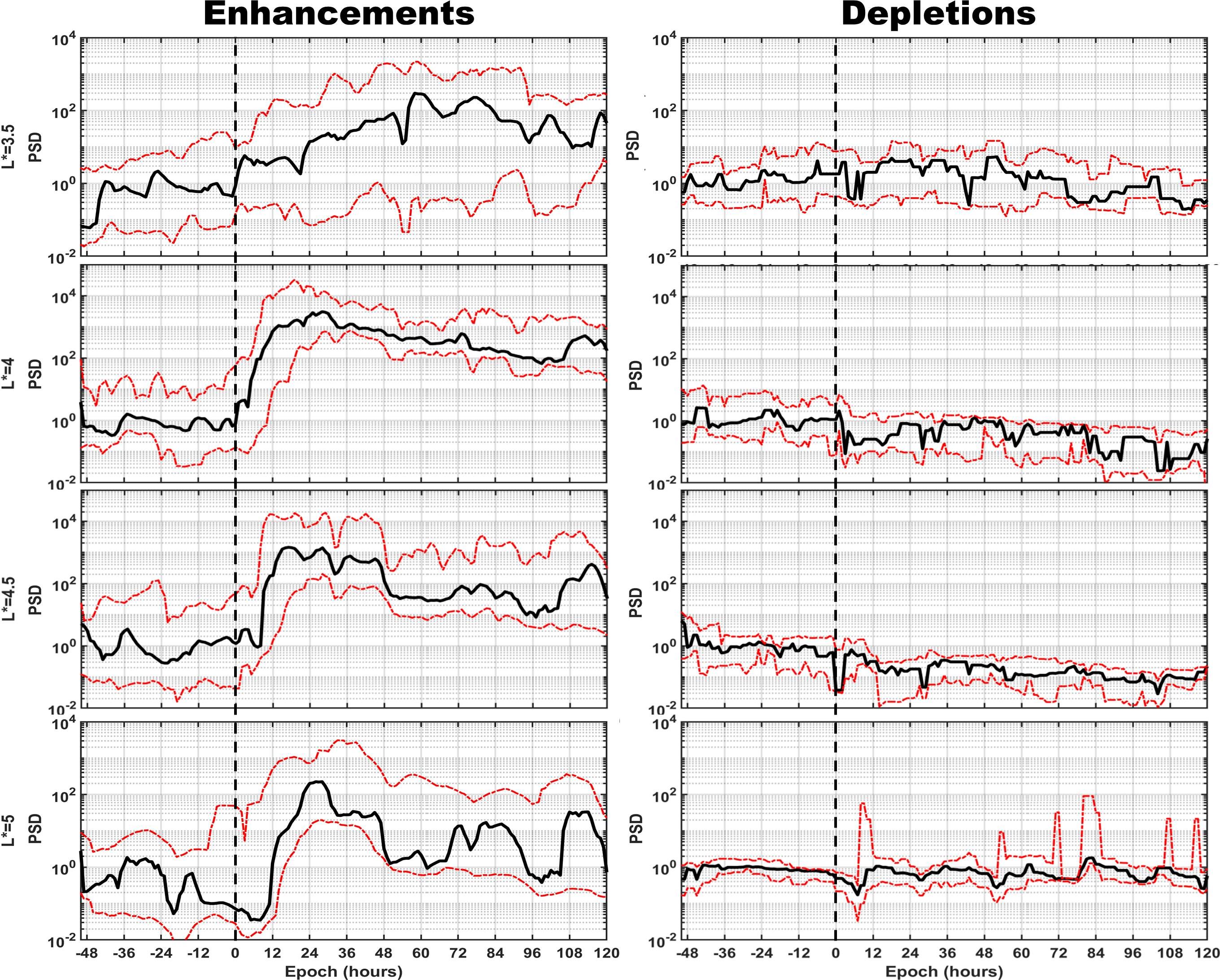}
 \caption{The same as figure \ref{ULF} for the normalized seed electron PSD with $\mu$=100 MeV/G.}
 \label{100}
  \end{figure}
%------------------------------------------------------------------------------------
Figure \ref{100} shows the SEA of the normalized PSD for the seed electron population ($\mu$=100 MeV/G) for 4 values for the third adiabatic invariant. As shown, the seed electron population, during enhancement events, exhibit clear increases up to 3 orders of magnitude, while during depletion events exhibit intermittent increases and decreases. This behaviour is consistent for all L$^{*}$ values. In detail, at L$^{*}$=4, enhancement events show a 3 orders of magnitude increase right after the maximum compression of the magnetopause which lasts until the end, while depletion events show a sudden 1 order of magnitude dropout, PSD is recovered until t$_{0}$+84 hours and, finally decreases again. At L$^{*}$=4.5 and 5, enhancement events present 2 to 3 orders of magnitude increase of the 100 MeV/G PSD, with maximum at 12 to 24 hours after the maximum compression of the magnetopause. On the contrary, depletion events, after a sudden dropout right after t$_{0}$, exhibit a gradual depletion at L$^{*}$=4.5 until the end of the post--phase and large variations at L$^{*}$=5. Note that, at L$^{*}$=3.5, enhancement events exhibit a gradual increase which reaches the maximum (more than 2 orders of magnitude) at approximately 60 hours after t$_{0}$. This behaviour indicates that -- except the enhanced substorm activity -- there must be additional mechanism(s) which produce this enhancement \citep{Turner2015b}. 

%------------------------------------------------------------------------------------
 \begin{figure}[h]
 \centering
 \includegraphics[width=38pc]{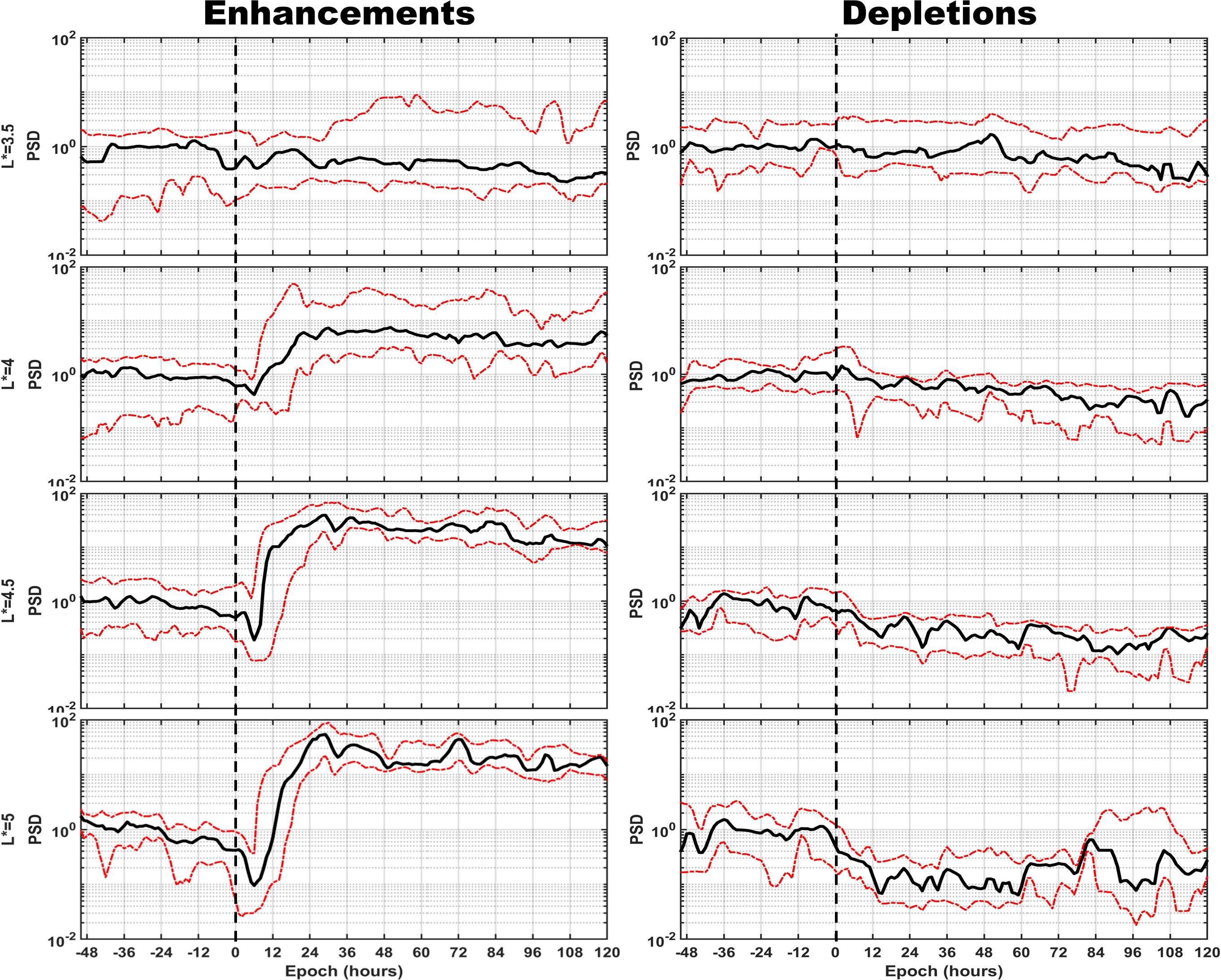}
 \caption{The same as figure \ref{ULF} for the normalized relativistic electron PSD with $\mu$=900 MeV/G.}
 \label{900}
  \end{figure}
%------------------------------------------------------------------------------------
Figure \ref{900} shows the SEA of the normalized PSD for the relativistic electron population ($\mu$=900 MeV/G) for 4 values of L$^{*}$. As shown, the electron PSD at L$^{*}$=3.5, exhibits slight decreases during both depletion and enhancement events. At higher L--shells (L$^{*}\geq$4.5), depletion events exhibit an -- up to 1 order of magnitude -- PSD decrease within 12 to 24 hours after t$_{0}$ at L$^{*}$=5 and 4.5, respectively. Then the population remains depleted until the end of the post--event phase. On the other hand, enhancement events exhibit a short--lived PSD dropout between t$_{0}$ and t$_{0}$+12 hours and then significant enhancements up to 2 orders of magnitude which remain until the end of the post--event phase. The fact that dropouts are more pronounced with increasing L$^{*}$ indicates that they are probably caused by magnetopause shadowing.
%------------------------------------------------------------------------------------
 \begin{figure}[h]
 \centering
 \includegraphics[width=38pc]{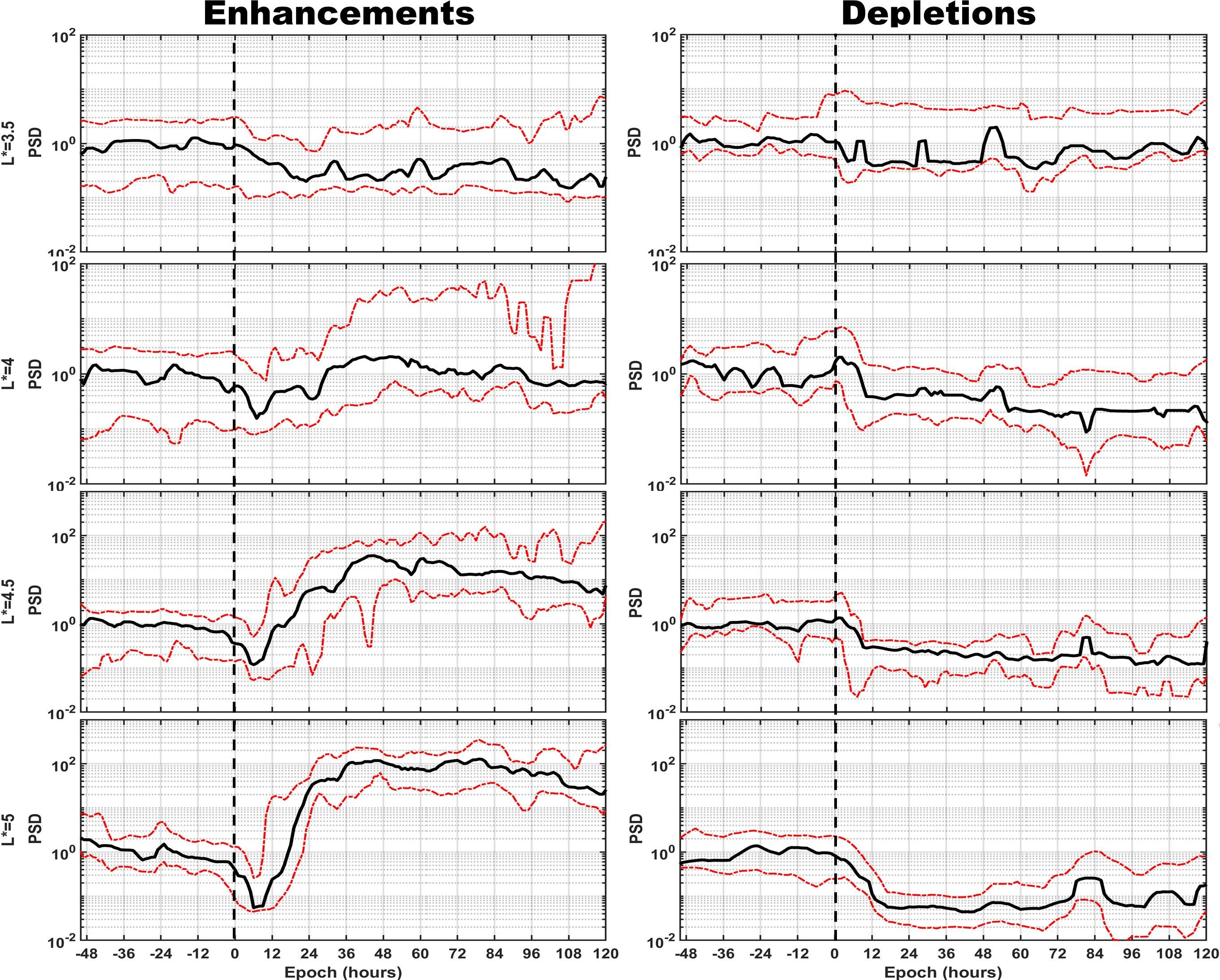}
 \caption{The same as figure \ref{ULF} for the normalized ultra--relativistic electron PSD with $\mu$=4200 MeV/G.}
 \label{4200}
  \end{figure}
%------------------------------------------------------------------------------------
Figure \ref{4200} shows the SEA of the normalized PSD for the ultra--relativistic electron population ($\mu$=4200 MeV/G) for 4 values of L$^{*}$. As shown, the electron PSD at L$^{*}$=3.5, exhibits the same slight decreases during both depletion and enhancement events, as the relativistic electrons. At higher L--shells (L$^{*}\geq$4.5), depletion events exhibit fast depletion within 12 hours after t$_{0}$ which remains until the end of the post--event phase, while at L$^{*}$=4, the decrease is more gradual. Enhancement events exhibit a short--lived dropout between t$_{0}$ and t$_{0}$+12 hours which is, again, more pronounced with increasing L$^{*}$. After the dropout, the PSD at L$^{*}$=4 returns to the pre--event phase levels. At higher L--shells (L$^{*}\geq$4.5), the PSD increases until it reaches the maximum (up to 2 orders of magnitude at L$^{*}$=5) between 36 and 48 hours after t$_{0}$.

\section{Discussion}
\subsection{Net effect statistics}

The net effect statistics show that the number of enhancements increases with increasing L$^{*}$ but decreases with increasing $\mu$. This is expected as fewer events have the ability of creating electrons that penetrate deep in the inner boundary of the outer belt and even fewer events can accelerate electrons to ultra--relativistic energies. This is consistent with the findings of  \citet{Reeves2016b} who showed that a given event is more likely to produce an enhancement of lower energy electron flux than it is to produce a higher energy enhancement. Moreover, enhancements of the seed electron population do not necessarily lead to relativistic electron enhancements. This is consistent with the results of \citet{Boyd2016} who showed that the seed population is subject to a threshold value that is a necessary condition for the enhancement of MeV electrons. Losses, on the other hand, appear to be L--dependent but not necessarily $\mu$--dependent. Finally, even though we cannot directly compare flux statistics with PSD, the relativistic electron (900 MeV/G) enhancement percentage coincides with previously reported results by \citet{Turner2015a}, but there is a significant difference concerning the depletion events (13\% in our study compared with 26\% in \citet{Turner2015a}) and those which had no significant change (47\% in our study compared with 35\% in \citet{Turner2015a}). A possible explanation lies in the fact we have selected events of which the majority lies during the declining phase of Solar cycle 24 which includes less intense activity than the maximum phase. Moreover, most of the statistical studies use a constant threshold (usually a factor of 2) to define enhancement/depletion instead of a sample-dependent threshold which is used here.

\subsection{Superposed epoch analysis}

From the initial sample of 71 events, we have selected 20 enhancement and 8 depletion events. Even though our sample is not large, it includes, almost equally, ICME and SIR driven events and is evenly spaced in the time period examined. Also note, the fact, that we have chosen those enhancements/depletions which are above the average enhancements/depletions of our sample. The latter gives us the advantage of having a clear view of the mechanisms occurring in each case. 

The application of SEA to solar wind parameters and geomagnetic indices, clearly shows that enhancement events are caused by geospace disturbances with persistently southward z component of the IMF combined with large and long--lasting values of solar wind speed. The strong reconnection which occurs due to the latter combination, leads to stronger and long--lasting decrease of SYM-H index, but most importantly, intense substorm activity as shown by the AL index. The importance of the intense series of substorm activity at the post--phase of enhancement events is shown in the seed population (100 MeV/G), especially at L$^{*}\geq$4, where this electron population exhibits an up to 3 orders of magnitude PSD increase. On the contrary, the same population exhibits mostly decreases at L$^{*}$=4 and 4.5, and intermittent variations at L$^{*}$=5. 

Moreover, substorm activity plays a fundamental role in radiation belt dynamics, since it is the cause of chorus waves generation due to anisotropic angular distributions of electrons with few to tens of keV's energy (typically referred as source population) which are injected near midnight from the plasma sheet due to substorm injections \citep{OBrien2003b,Baker2007,Thorne2013}. Thus, chorus and substorm activity are strongly associated \citep{Meredith2001,Li2015,Boynton2018}. On the other hand, depletion events are accompanied by weak and short--lived southward z component around t$_{0}$ which quickly turns northward up to 2 days after t$_{0}$. Combined with lower values of solar wind speed leads to a rather weak and short--lived substorm activity as indicated by the AL index. All of the above are in good agreement with the findings by \citet{Li2015} who showed that chorus wave activity is more pronounced and long--lived, over a broad L--shell region during enhancement events. The latter is verified by the results of this work, as chorus amplitudes, during enhancement events, exhibit stronger (up to a factor of 80) and more long--lived enhancements (up to 4 days after t$_{0}$) at L$^{*}\geq$4.5, compared to depletion events.
 
Pc5 activity, on the other hand, does not exhibit significant difference concerning the maximum power between enhancement and depletion events but exhibits significant differences concerning the duration of enhanced activity (3.5 and 1.5 days after t$_{0}$, respectively). \citet{Turner2013}, by comparing two storms in detail, showed that the one that resulted in enhancement -- compared to the one that resulted in depletion of relativistic electron PSD -- exhibited more enhanced and broader range chorus wave amplitudes as well as, more prolonged periods of enhanced Pc5 wave activity. Our statistical results, verify the aforementioned distinction between the two groups of events by using a much larger sample.

Our results indicate that the combined effect of magnetopause shadowing and outward diffusion driven by Pc5 waves is present in both groups of events. This is expected since the maximum compression of the magnetopause and the minimum of Bz are comparable between enhancements and depletions and, in addition, differences in Pc5 activity are quite small within 12 to 24 hours around t$_{0}$. The fact that loss of radiation belt electrons via magnetopause shadowing is a common feature during the initial phase of storms, is in agreement with the findings of \citet{Murphy2018}, yet we add that such feature is also present in weak or even non--storm events. Further support, to the aforementioned indication, is provided by the fact that these losses are more pronounced with increasing values of both $\mu$ and L$^{*}$. 

Moreover, the PSD presented in this work, corresponds to near--equatorial mirroring electrons with equatorial pitch angles, approximately within the 70--90 degrees range, which are not directly affected by EMIC waves \citep{Usanova2014}.

All of the above, support the scenario that, in the case of enhancement events, the existence of enhanced seed population and chorus activity can quickly replenish the losses of relativistic electrons due to combined magnetopause shadowing and outward diffusion, while inward diffusion can further accelerate them to higher energies. This is consistent with the dual role of ULF--Pc5 waves, not only in losses via outward diffusion but also in acceleration of relativistic electrons to ultra--relativistic energies \citep{Jaynes2015}. 

On the contrary, in the case of depletion events, the absence of enhanced seed population and chorus activity, renders the  combination of magnetopause shadowing and outward diffusion, as the dominant loss mechanism.

\section{Conclusions}
We performed a statistical analysis of 71 geospace disturbances that occurred during the RBSP era (September 2012 -- April 2018). Our results show that the number of enhancements increases with increasing L$^{*}$ but decreases with increasing $\mu$. Depletions on the other hand, appear to be L--dependent but not necessarily $\mu$--dependent. 

The application of superposed epoch analysis to the 20 enhancement and 8 depletion events which are above the average enhancements/depletions has shown that:

\begin{enumerate}

\item Enhancement events are caused by geospace disturbances with persistently southward z component of the IMF combined with large and long--lasting values of solar wind speed.

\item Consequently, the enhanced reconnection, leads to stronger and long--lasting decrease of SYM-H index, but most importantly, leads to intense substorm activity -- as indicated by the AL index -- and thus, more intense and long--lived chorus activity.

\item Due to the substorm activity, enhancement events exhibit increase of the seed electron PSD, more than 3 orders of magnitude, while depletion events exhibit weaker increases less than 2 orders of magnitude.

\item Pc5 activity does not exhibit significant difference concerning the maximum power, but exhibits significant differences concerning the duration of enhanced activity, up to 3.5 and 1.5 days after t$_{0}$ for enhancement and depletion, respectively.

\item During enhancement events, the existence of enhanced seed population and chorus activity can quickly replenish the losses of relativistic electrons due to combined magnetopause shadowing and outward diffusion.

\item During depletion events, the absence of enhanced seed population and chorus activity, renders the combination of magnetopause shadowing and outward diffusion, as the dominant loss mechanism. 

\end{enumerate} 

% ------------------------------------------------------------------------ %
%  ACKNOWLEDGMENTS
\acknowledgments
The authors acknowledge the RBSP/MagEIS and RBSP/REPT teams for the use of the corresponding datasets. MagEIS (release 4) and REPT (release 3) data are available in https://www.rbsp-ect.lanl.gov/science/DataDirectories.php. Wen Li would like to acknowledge the NASA grant NNX17AD15G, the AFOSR grant FA9550-15-1-0158, the NSF grant AGS-1723588, and the Alfred P. Sloan Research Fellowship FG-2018-10936. We acknowledge the NASA CDAWeb for providing SYM-H index and solar wind parameters’ data and the POES/SEM team for the use of the corresponding dataset (https://satdat.ngdc.noaa.gov/sem/poes/data/).
% ------------------------------------------------------------------------ %

\end{document}